\documentclass[twocolumn]{aastex631}
\shorttitle{First stars in FDM cosmology}
\shortauthors{Kulkarni et al.}
\graphicspath{{./}{figures/}}
\usepackage{bm}
\usepackage{physics}
\usepackage{savesym}
\savesymbol{tablenum}
\usepackage{siunitx}
\restoresymbol{SIX}{tablenum}
\begin{document}

\title{If dark matter is fuzzy, the first stars form in massive pancakes}

\author[0000-0002-9789-6653]{Mihir Kulkarni}
\affiliation{Department of Physics and Astronomy and Ritter Astrophysical
Research Center, University of Toledo, 2801 W Bancroft Street, Toledo, OH 43606, USA}

\author[0000-0002-8365-0337]{Eli Visbal}
\affiliation{Department of Physics and Astronomy and Ritter Astrophysical
Research Center, University of Toledo, 2801 W Bancroft Street, Toledo, OH 43606, USA}

\author[0000-0003-2630-9228]{Greg L. Bryan}
\affiliation{Department of Astronomy, Columbia University, 550 W 120th Street, New York, NY 10027, USA}

\author[0000-0003-0750-3543]{Xinyu Li}
\affiliation{Canadian Institute for Theoretical Astrophysics, 60 St George St, Toronto, ON M5R 2M8, Canada}



\begin{abstract}

Fuzzy dark matter (FDM) is a proposed modification 
for the standard cold dark matter (CDM) model motivated by small-scale discrepancies in low-mass galaxies. Composed of ultra-light (mass $\sim \SI{e-22}{eV}$) axions with kpc-scale de Broglie wavelengths, this is one of a class of candidates that predicts that the first collapsed objects form in relatively massive dark matter halos. This implies that the formation history of the first stars and galaxies would be very different, potentially placing strong constraints on such models. Here we numerically simulate the formation of the first stars in an FDM cosmology, following the collapse in a representative volume all the way down to primordial protostar formation including a primordial non-equilibrium chemical network and cooling for the first time. We find two novel results: first, the large-scale collapse results in a very thin and flat gas “pancake”; second, despite the very different cosmology, this pancake fragments until it forms protostellar objects indistinguishable from those in CDM. Combined, these results indicate that the first generation of stars in this model are also likely to be massive and, because of the sheet morphology, do not self-regulate, resulting in a massive Pop III starburst. We estimate the total number of first stars forming in this extended structure to be $10^4$ over 20 Myr using a simple model to account for the ionizing feedback from the stars, and should be observable with JWST. These predictions provide a potential smoking gun signature of FDM and similar dark matter candidates. 

\end{abstract}


\submitjournal{ApJL}
\accepted{November 21, 2022}

\section{Introduction} \label{sec:intro}

The standard model of cosmology, which includes dark energy in the form of a cosmological constant and `cold' dark matter, has been extremely successful explaining the large-scale structure in the universe such as the power spectrum of the cosmic microwave background, cluster abundances, and galaxy clustering \citep{Colberg2000, Mo02, Bennett13, Planck18}. However, it faces some apparent problems on small scales such as missing satellite problem \citep{Klypin99}, too-big-to-fail problem \citep{Boylan11_toobigtofail}, cusp-core problem \citep{NFW, Burkert95, Goerdt2006}. A number of solutions have been proposed to alleviate these problems including strong baryonic feedback processes \citep[e.g.][]{Navarro96, Onorbe15, GarrisonKimmel19}. Additionally, a number of alternative dark matter models, such as the fuzzy dark matter, warm dark matter, self-interacting dark matter, have also been proposed over the years \citep{Hu2000_fdm, Bode01, Tulin18}. They primarily differ from standard cold dark matter on small spatial scales, typically having a sharp cut-off in the matter power spectrum for wave numbers above a characteristic scale.

Fuzzy dark matter (FDM), made up of ultra-light axions, is theorized to have a particle mass of $\sim 10^{-22}$ eV to make it astrophysically relevant for the small-scale problems \citep{Hu2000_fdm, Marsh16_review, Hui17}. Its extremely small mass makes the associated de Broglie wavelength of the order of a kpc, relevant to astrophysical scales. This results in an effective ``quantum pressure'', suppressing small scale structures. Thus, fuzzy dark matter cosmology has a small-scale cutoff in the matter power spectrum which results in the suppression of dark matter halos with mass lower than \SI{e9}{M_\odot} \citep{Schive14_nature, Schive16_simulation, Kulkarni22_fdm, May22}. 

A number of constraints have been placed on the properties of the fuzzy dark matter using their predicted observed astrophysical signatures. The velocity dispersion of stars in dwarf spheroidal galaxies has been used to infer the size of the cores in dark matter density profiles, constraining the axion mass \citep{Calabrese16}. Based on the dynamics of the stellar streams in the Milky Way the axion mass has been constrained \citep{Amorisco18}. Some of the strongest constrains on the axion mass come from the Lyman-$\alpha$ forest \citep{Armengaud17, Irsic17, Kobayashi17}. 

In this letter, we simulate the formation of the first stars and galaxies in a fuzzy dark matter cosmology, which can be used to put strong constraints on the mass of axion. In the $\Lambda$CDM cosmology, numerical simulations predict that the first stars, also known as Population III (Pop III) stars typically first form in minihalos of mass $10^5 - $\SI{e7}{M_{\odot}}, when the gas can cool via rotational and vibrational transitions of molecular hydrogen \citep{Haiman96, Tegmark97, Abel2002, Yoshida2003, Oshea07, Wise-Abel07, Kulkarni21}. In the absence of minihalos, the formation of first stars and galaxies is significantly delayed in a fuzzy dark matter cosmology and are thus expected to form in much more massive dark matter structures at much lower redshifts, resulting in a very different star-formation history at high redshift. This places constraints on the properties of the fuzzy dark matter and other dark matter models that predict sharp small-scale cutoffs.

Previously a few groups have addressed this problem with approximate methods or in other models. \cite{Gao07} studied this question for a warm dark matter model and found that the first stars form in dark matter filaments. \cite{Hirano18} studied the formation of first star-forming structures in fuzzy dark matter cosmology using an N-body collisionless code with the FDM power spectrum. \cite{Mocz19,Mocz20} studied this problem for the first time while accurately evolving the Schr{\"o}dinger-Poisson equations; their star-formation criteria does not specifically account for the cooling processes at low or zero metallicity and so was not designed to predict the formation of the first stars in primordial gas. In this work, for the first time, we accurately evolve the FDM density by solving the Schr{\"o}dinger-Poisson equations and follow the protostellar collapse of gas with a non-equilibrium primordial chemistry network and radiative cooling with an adaptive mesh refinement at high resolution. 

We find that the first stars form in a sheet-like dark matter structure reminiscent of a Zel'dovich pancake \citep{Zeldovich70}. This geometry results in a burst of Pop III stars with minimal feedback effects resulting in the production of a stellar mass of $10^5 - $ \SI{e6}{M_{\odot}} over a time of approximately 20 Myr. The geometry results in dense gas spread over a sheet resulting in a large number of star-forming clumps that are also spread out over a larger spatial scale resulting in lowered feedback effects.
Observing such a massive Pop III starburst would be a smoking gun signature for FDM or other dark matter models, which could be detected with the recently launched James Webb Space Telescope (JWST).

This letter is structured as follows. In Section~\ref{sec:methods}, we describe the numerical methods we used to generate the initial conditions and to accurately evolve the dark matter and gas distribution. In Section~\ref{sec:results}, we explain our results about the sheet geometry, the collapse of protostars, and the distribution of other clumps in the sheet. In Section \ref{sec:discussion}, we discuss the implications for our results, the simple feedback prescription used to estimate the total number Pop III stars, as well as the observational prospects for JWST. We then summarize our results and main conclusions in Section \ref{sec: conclusions}.

\section{Methods}
\label{sec:methods}
We perform our cosmological simulation using the adaptive mesh refinement (AMR) code \textsc{Enzo} \citep{Enzo, Enzo_joss}. We use the energy conserving, spatially third-order accurate Piecewise Parabolic Method for
the hydro solver. \textsc{Enzo} follows the non-equilibrium evolution of nine species (H, H$^+$, He, He$^+$, He$^{++}$, e$^{-}$, H$_2$ , H${_2^+}$ , and H$^{-}$). We used cosmological parameters from \cite{Planck18}: $H_0$ = \SI{67.36}{km/s/Mpc}, $\Omega_m$ = 0.315, $\Omega_b$ = 0.0493, $n_s$ = 0.9649. We change $\sigma_8$ from 0.811 to 1.4 as described later in this section.

To accurately follow the evolution of the fuzzy dark matter distribution, we solve the Schr{\"o}dinger-Poisson equations on a uniform grid using the method described in \citet{XinyuLi19}. The wavefunction $\psi$ evolves according to the Schr{\"o}dinger equation as 
\begin{equation}
    i \hbar \left(\partial_t \psi + \frac{3}{2} H \psi \right) = \left(-\frac{\hbar^2}{2 m_a a^2} \nabla^2 + m_a \Phi \right) \psi, 
\end{equation}
where $t$ is the cosmic time, the spatial derivative is with respect to the comoving coordinates, $m_a$ is the axion mass, $a$ is the scale factor, $H$ is the Hubble parameter and $\Phi$ is the gravitational potential. The gravitational potential is calculated as
\begin{equation}
    \nabla^2 \Phi = 4 \pi G a^2 (\rho - \bar{\rho}),
\end{equation}
where $\bar{\rho}$ is the cosmic mean mass density and include both FDM and baryonic contributions ($\rho = \rho_{\rm FDM} + \rho_b$). The wave function relates to the FDM density as 
\begin{equation}
    \psi \equiv \sqrt{\frac{\rho_{\rm FDM}}{m_a}} e^{i \theta}; \quad \rho_{\rm FDM} = m_a \abs{\psi}^2.
    \label{eq:psi_rho}
\end{equation}
The velocity field for the dark matter is related to the phase of the wavefunction as
\begin{equation}
    \bm{v} \equiv \frac{\hbar}{m_a a} \nabla \theta.
    \label{eq:v_fdm}
\end{equation}

To generate the initial conditions, we use a modified version of MUSIC \citep{Music} with a power spectrum for FDM given as
\begin{equation}
    P_{FDM}(k) = T_F^2(k)  P_{CDM}(k),
\end{equation}
where $P_{CDM}(k)$ is the CDM power spectrum, $P_{FDM}(k)$ is the FDM power spectrum and $T_F(k)$ is the FDM transfer function
\begin{equation}
    T_F = \frac{\cos x^3}{1 + x^8}
\end{equation}
as described in \cite{Hu2000_fdm}, 
where $x = 1.61 m_{22}^{1/18} k/k_{Jeq}$. $k_{Jeq}$ is the Jeans length at the matter-radiation equality given as $k_{Jeq} = 9 m_{22}^{1/2}\si{Mpc^{-1}}$. Here $m_{22} = m_a / (\SI{e-22}{eV})$. We generate the FDM density and velocity fields using MUSIC at $z = 100$. We use Eq. \ref{eq:psi_rho} and \ref{eq:v_fdm} to calculate the real and imaginary parts of the wavefunction $\psi (\bm{x})$ to be used by \textsc{Enzo} as the initial condition to solve the Schr{\"o}dinger equation.

For our simulation, we use a box size of \SI{1.7}{h^{-1}Mpc} with $m_a = \SI{2.5e-22}{eV}$. This corresponds to a half-mode wavelength of approximately 800 comoving kpc. To ensure that we have sufficiently collapsed structure in this small box, we increase the amplitude of DM perturbations by increasing $\sigma_8$ to $1.4$ instead of $0.811$ given by \cite{Planck18} implying that we are effectively simulating an overdense region. 
We evolve the FDM wave function on a uniform grid of $1024^3$. This ensures that the de Broglie wavelength of the FDM is resolved everywhere in the box. We set the time step constraint as described in \cite{XinyuLi19}.

To accurately follow the evolution of star-forming clumps with high spatial resolution, we add the AMR for the gas evolution in a refinement region of size \SI{0.119}{h^{-1} Mpc} comoving (0.07 times the box size) at $z = 13$ centered at the densest region in the box. The density evolution before this time is quasi-linear. The cells are refined when the gas density in a cell reaches above $4\times 2^{3l}$ times the background gas density on the root grid, where $l$ is the refinement level. We also employ a refinement criterion based on the Jeans length such that the local Jeans length is always resolved by at least 16 cells. We evolve the dark matter density only on the root grid and interpolate it on the fine grids. To calculate the gravitational field, we included mass from the baryons in the refined cells and the interpolated DM mass. This allows us to follow the baryon-dominated protostellar collapse to very high density in the regions of interest. We use a total of 9 levels of refinement corresponding to a resolution of $\sim 0.4$ pc (proper) at $z = 10$. We stopped our simulation when it reaches 9 levels of refinements as it became prohibitively expensive to continue the simulation further. The gas mass dominates over the dark matter mass in the refined cells on small scales --- the most refined cell has a gas density $\sim 10^5$ times the dark matter density, thus suggesting that our approximate treatment for the dark matter on the root grid is accurate for studying the protostellar collapse.

\section{Results}
\label{sec:results}
In this section we present our results. We first describe the geometry of the sheet at the time of star formation, before turning to the properties of the first collapsing protostar. Finally, we describe how we identify other gas clumps in the sheet and assess their fate.

\subsection{Geometry of the sheet}
\begin{figure*}
    \centering
    \includegraphics[width=0.95\textwidth]{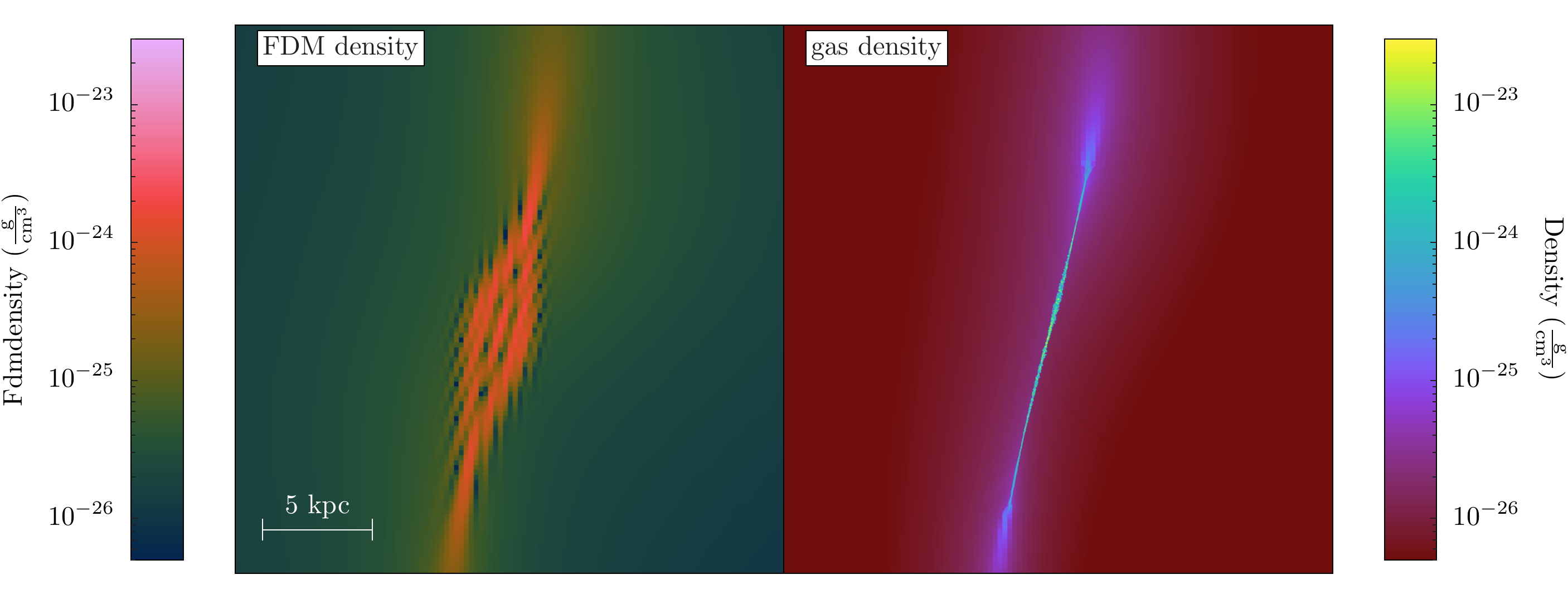}
    \caption{The dark matter density (left) and the gas density (right) slices at $z = 10.03$ passing through the sheet. FDM density shows the characteristic interference pattern. The gas sheet collapses on scales much smaller than the FDM structure.}
    \label{fig:fdm_slice}
\end{figure*}

We find that when the first protostar in our simulation collapses to a number density of \SI{2e5}{cm^{-3}} (at $z = 10.03$), it is located in a large Zel'dovich pancake-like structure instead of a quasi-spherical dark matter halo (as expected in $\Lambda$CDM). The sheet extends over $15-20$ kpc in the plane. The thickness of the FDM sheet is $\sim$ 2 kpc, whereas the gas structure is thinner and has a thickness of $\sim$ 200 pc. Figure~\ref{fig:fdm_slice} shows slices of FDM and gas density at the point of runaway collapse. As the dark matter structure collapses along the first eigenvector, it forms a two-dimensional sheet (with interference fringes in the fuzzy case). This general kind of collapse has been studied in the past, particularly in the context of hot dark matter \citep[e.g.][]{Zeldovich70, Anninos94, Anninos95}. In the presence of the small-scale power in a CDM cosmology, instead of this large sheet of gas, there would be many small collapsed halos in the plane that are absent in the fuzzy dark matter cosmology. Therefore, we expect a similar geometry during the formation of first stars for all dark matter models that have a small-scale cutoff in the power spectrum. \cite{Gao07} simulated the formation of first stars in warm dark matter cosmology and found that the first stars form along a dark matter filament of size \SI{3}{kpc} instead of a sheet. \cite{Gao07} used WDM with a mass of \SI{3}{keV} that suppresses power below the scales of \SI{100}{kpc} as compared to our axion mass of \SI{2.5e-22}{eV} that suppresses power over much larger scales of approximately \SI{800}{kpc}. As our choice of dark matter mass is ``warmer"/``fuzzier" (i.e. a larger cutoff), this suggests that the dark matter collapse follows a \emph{sheet} $\rightarrow$ \emph{filament} path as the dark matter gets less fuzzy, and that the exact dark matter structure geometry when the first stars form depends on the particle mass and formation epoch.

\begin{figure*}
    \centering
    \includegraphics[width=0.45\textwidth]{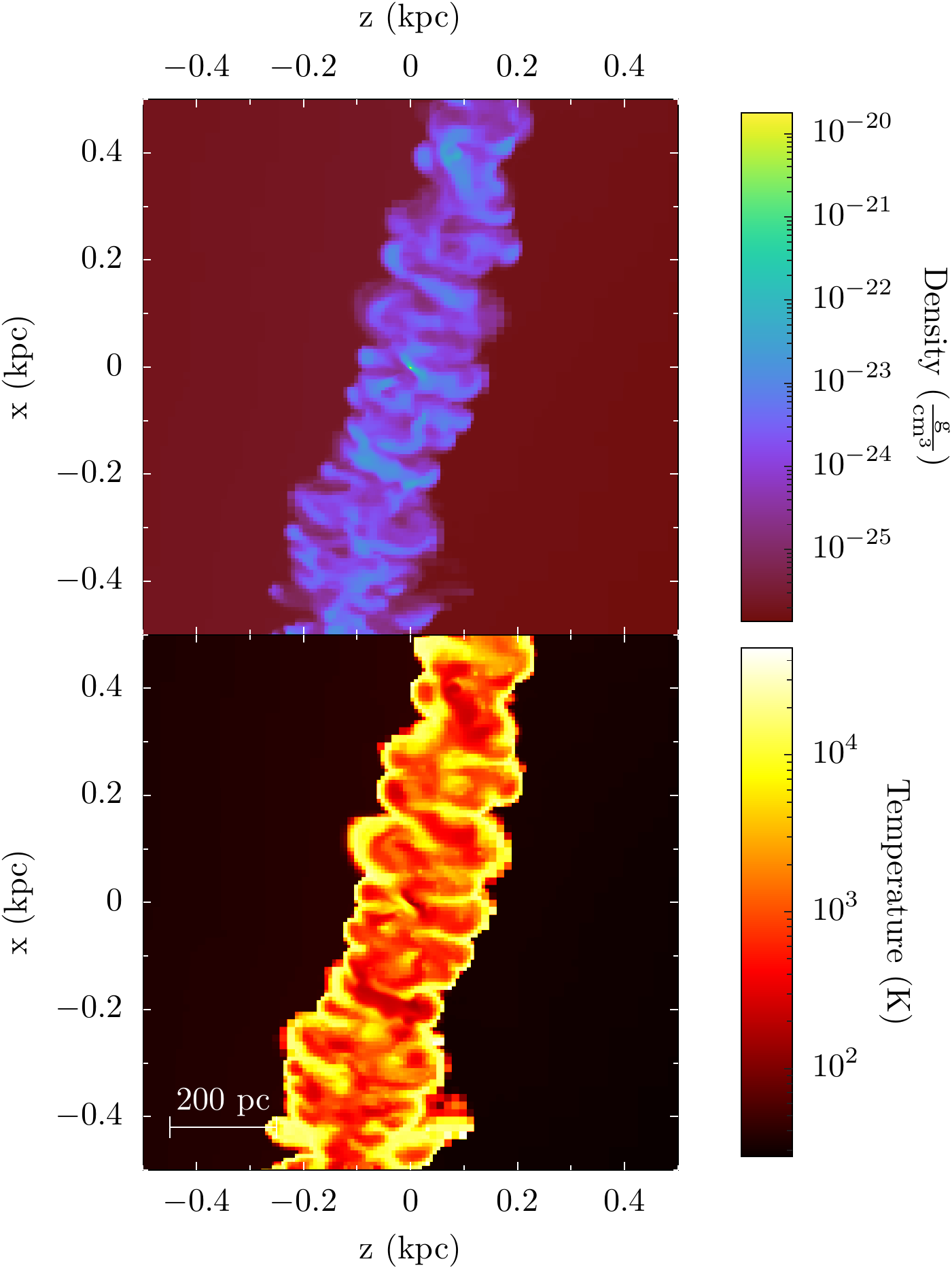}
    \includegraphics[width=0.4\textwidth]{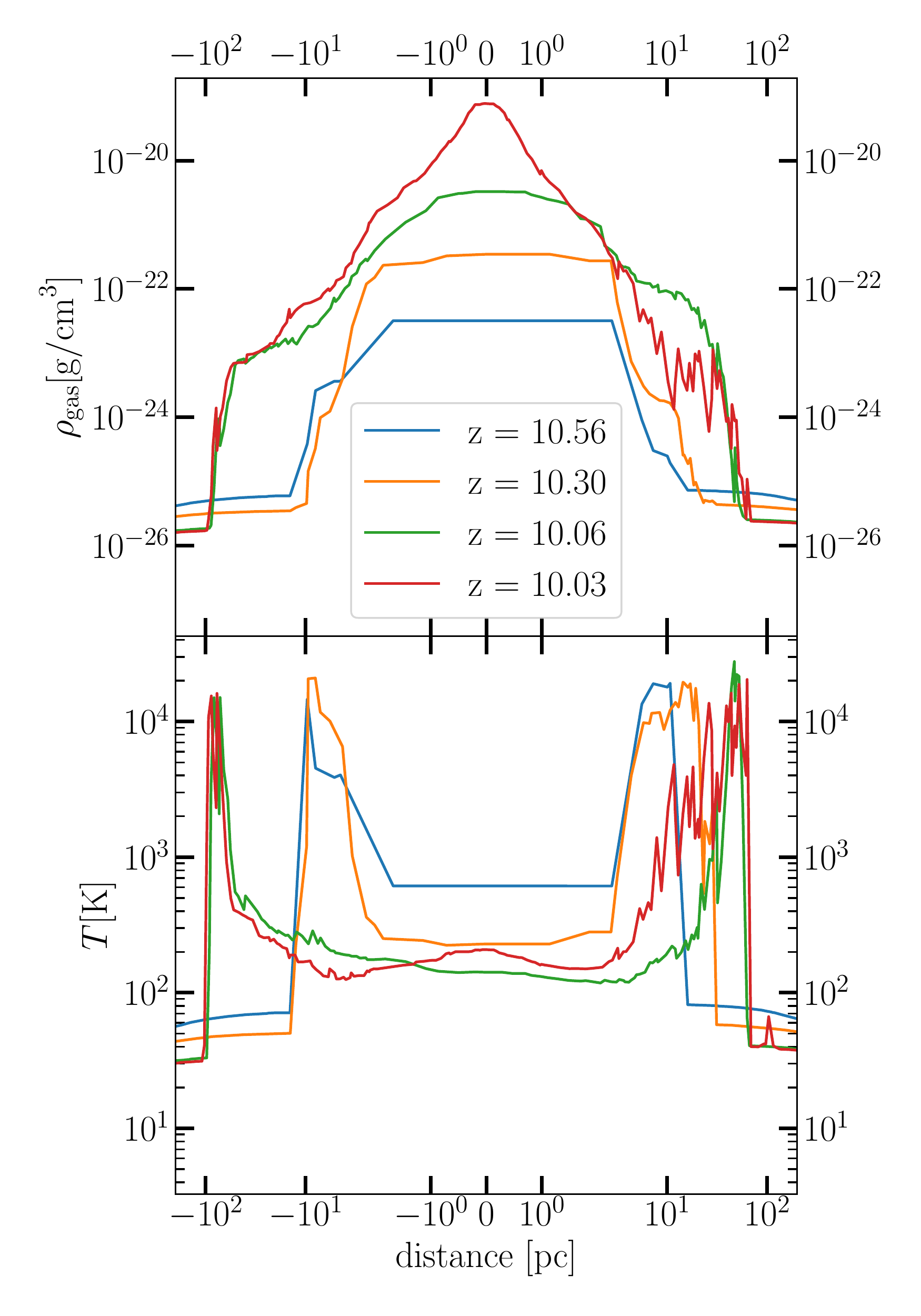}
    \caption{The left panels show slices of gas density and gas temperature in a plane passing through the pancake centered at the densest cell at $z = 10.03$. The gas density is highest in the mid-plane, where the gas temperature is reduced to a temperature of $200-300$ K by molecular hydrogen cooling. The gas on the edge of the sheet is shock-heated to temperatures of $10^4-10^5$ K. The right panels show the gas density and temperature profiles for a ray passing through the densest cell perpendicular to the sheet as a function of time. The width of the sheet increases with time as more gas falls onto it and the mid-plane gets denser with time. The temperature profile shows a M-shape, where the gas is shock-heated near the edge and is cooled in the center. The thickness of the sheet at $z = 10.03$ approximately 200 pc.} 
    \label{fig:sheet_geometry}
\end{figure*}

Figure \ref{fig:sheet_geometry} shows the gas density and temperature slices at $z=10.03$ in a plane passing through the sheet (left panels), and the gas and temperature profiles along a ray perpendicular to the sheet passing through the densest clump as a function of time (right panels). 
Going from outside in, the gas density increases at the edge of the sheet, reaches a plateau inside, before increasing rapidly again in the central region where the gas is rapidly cooling and collapsing to higher densities. On the other hand, near the edge of the sheet, the gas temperature rises to $10^4 -$  \SI{e5}{K} because of the shock-heated gas infalling from the perpendicular direction. In the inner region of the sheet, the gas is able to cool from the rotational and vibrational transitions of molecular hydrogen, to temperatures as low as $\sim \SI{200}{K}$ in the central collapsing clump. This results in most of the clumps in the pancake being near the center of the plane. For a detailed discussion on pancake collapse, see \cite{Anninos94, Anninos95}.

The infalling gas that heats up the edges of the sheet also adds to the size and mass of the sheet in time. The right panels show the gas density and temperature profiles perpendicular to the sheet, which shows a generally increasing thickness of the sheet.

\begin{figure}
    \centering
    \includegraphics[width=\columnwidth]{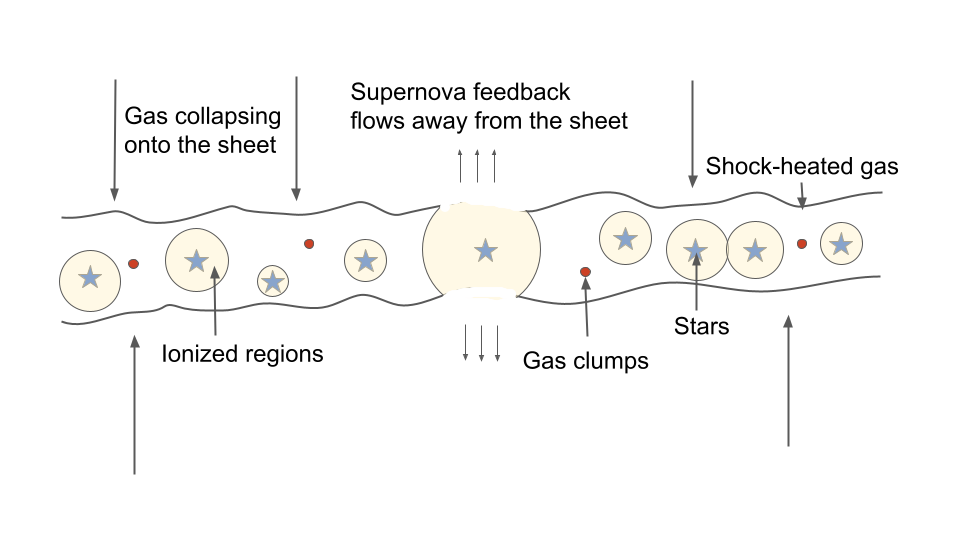}
    \caption{Schematic diagram showing the edge-on view of the sheet and our expectation of its continued development. The gas falls onto the sheet from a perpendicular direction that is shock-heated when it reaches the the sheet. The red dots denote the dense gas clumps that haven't turned into stars yet. Blue stars and yellow circles around them show the stars that have formed and the ionized regions around them that grow with time. When an ionized region reaches the top or bottom of the sheet, the ionization and supernovae feedback escape away from the sheet in the perpendicular direction. This minimizes the ionization and supernovae feedback effects on subsequent star formation within the sheet.}
    \label{fig:schematic}
\end{figure}

Figure \ref{fig:schematic} shows a schematic diagram describing the sheet geometry from an edge-on perspective, as well as our expectation of its evolution beyond the formation of the first clump. The dense star-forming clumps are spread out near the central plane of the sheet and turn into stars at different times depending on their densities. The ionizing and supernovae feedback from these stars create bubbles that prevent subsequent star-formation in them. The details of the feedback processes are discussed in Section~\ref{sec:discussion}.

\subsection{Collapse of the central protostar}
\begin{figure}
    \centering
    \includegraphics[width=\columnwidth]{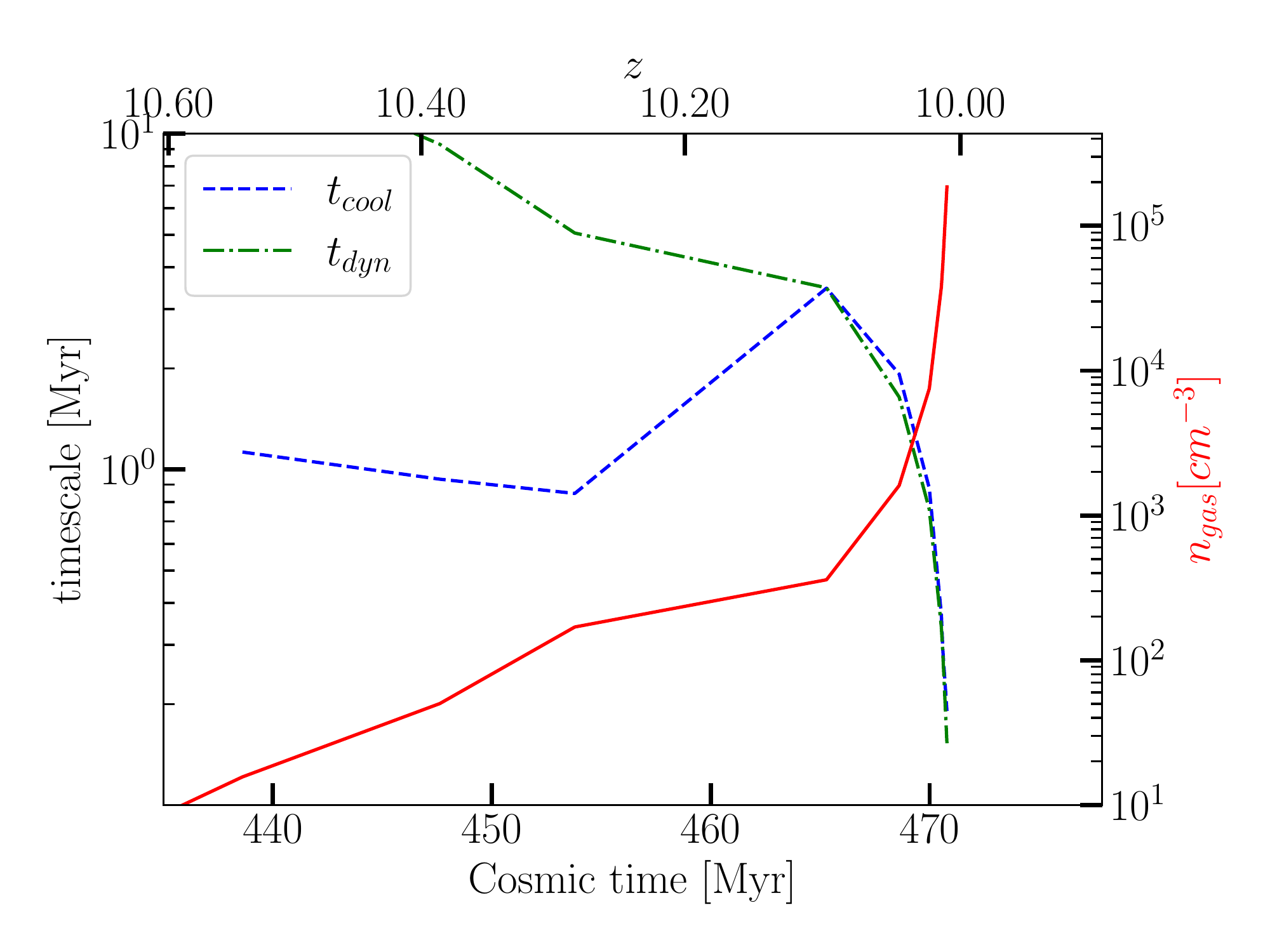}
    \caption{Runaway collapse of the central protostar. The $y$-axis on the left shows the timescales (dynamical time in green dot-dashes and cooling time in  blue dash) as a function of time for the central protostar in the sheet. Initially the dynamical time is large as the densities are small. As the runaway collapse starts near $t = 465 ~{\rm Myr}$, the cooling time and the dynamical time decrease rapidly and follow each other.  This results in the central gas density rapidly increasing (red solid curve) displaying the collapse.}
    \label{fig:protostar}
\end{figure}

We follow the runaway collapse of the central densest protostar in the sheet using adaptive mesh refinement. The highest refinement level reached for the central protostars is 9, which results in the highest resolution of $\sim \SI{0.4}{pc}$ (proper). Figure~\ref{fig:protostar} shows the evolution of the cooling time, dynamical time and the gas density for the central protostar as a function of time. The blue dashed line and the green dot-dashed line denote the cooling time and the dynamical time of the densest cell in the clump as a function of time. Prior to a cosmic time of \SI{465}{Myr}, the clump density is significantly lower resulting in the dynamical time being longer than the cooling time. After \SI{465}{Myr}, the cooling time and the dynamical time nearly follow each other and decrease rapidly. During this period, the gas density increases rapidly because of efficient cooling as shown by the red solid line.

\begin{figure}
    \centering
    \includegraphics[width=\columnwidth]{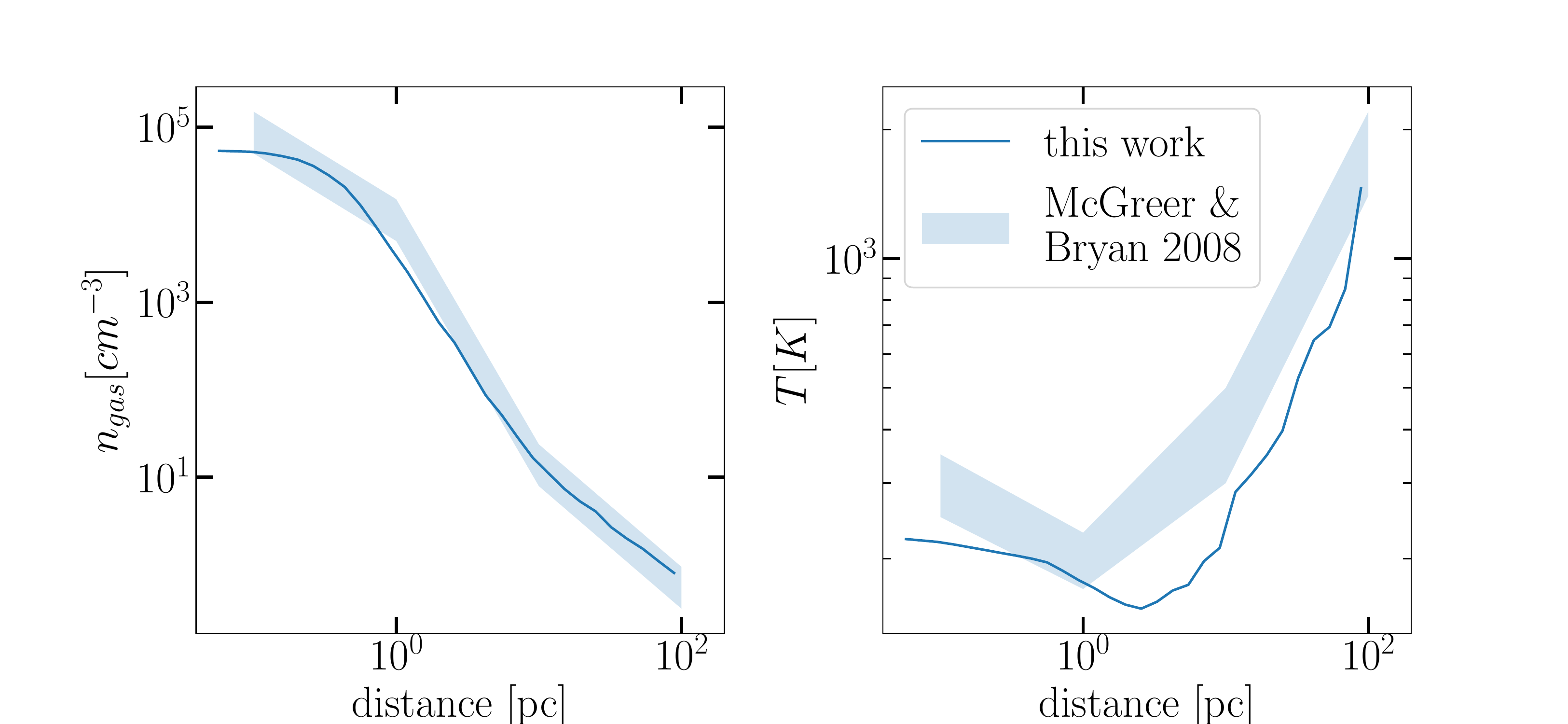}
    \caption{The density (left) and the temperature profile (right) of the most massive clump (solid). The shaded region shows the radial profiles around four clumps with H$_2$ cooling from \cite{McGreer08} in $\Lambda$CDM minihalos suggesting that the properties of the first stars in FDM cosmology are similar to those in $\Lambda$CDM cosmology.}
    \label{fig:clump_profile}
\end{figure}

We find that the density evolution of the central protostellar clump is similar to the evolution of Pop III star-forming clumps in minihalos as described in \cite[e.g.,][]{Abel2002}. Figure~\ref{fig:clump_profile} shows the radial density and temperature profiles around the most massive clump. The shaded region shows the profiles from \cite{McGreer08} around the collapsed clumps in $\Lambda$CDM minihalos. These are consistent with the profiles we see here and therefore we conclude that the properties of the first stars forming in the pancakes in the FDM cosmology are similar to ones forming in minihalos in $\Lambda$CDM, even though the underlying dark matter distribution is very different. This differs somewhat from the arguments made in \cite{Gao07} suggesting a lower mass for stars forming in WDM cosmologies in the absence of small scale perturbations that may trigger fragmentation at low gas densities. In contrast, we find that the formation of the shock with the associated shear flow may allow for the development of fluid instabilities on small scales, seeding collapse on multiple scales and resulting in star-formation similar to $\Lambda$CDM minihalos.
This remarkable result shows the robustness of the molecular-hydrogen mediated, baryon-dominated collapse that leads to Pop III stars.

\subsection{Other clumps}
When we stop the simulation at $z = 10.03$, the gas is distributed in a sheet as described in the previous subsections. The gas in the sheet is fragmented into multiple clumps that are potentially star-forming locations. Although we do not simulate their collapse, we expect each self-gravitating clump to also follow the Pop III attractor solution. We identify these clumps using a simple procedure that depends only a characteristic distance $l$ (set as a free parameter), as follows. We first arrange all the cells with number densities higher than $n = \SI{10}{cm^{-3}}$ by decreasing gas densities. We then identify a cell as a clump if it is more than a distance $l$ away from all the clumps denser than itself. If a cell is within a distance $l$ of one of the clumps already identified, it is assigned to that clump. Thus, we have densities, locations and gas masses associated with independent clumps that are separated from each other by at least a distance $l$. We changed the distance parameter $l$ to be 5, 10, 15, and 30 pc. The number of clumps decrease with increasing $l$ since some of the clumps that were identified as separate clumps become a single clump as the distance parameter increases. We find that the number of clumps is converged for $l=15$ pc, so we use that value in our calculations. We find about 50,000 clumps in the sheet at the time when the simulation undergoes runaway collapse in the densest clump (and the simulation is stopped). To identify which of the clumps collapse to form stars in the presence of feedback effects from other stars, we implement a simple analytic model, which we describe in the next section.

\section{Discussion}
\label{sec:discussion}

\subsection{Ionization feedback}
As the first stars form, they start emitting ionizing radiation. This creates ionized bubbles that have $d$-type shock fronts around them. This generally destroys any star-forming clumps in the ionized regions and prevent their formation into stars. When the stars die, they can explode as supernovae depending on their mass. This enriches nearby gas with metals and has an associated supernova shock. When this metal-enriched gas cools it results into formation of next generation Pop II stars. When the ionized and supernovae shocked regions reach the edge of the sheet, they encounter a region of low pressure and thus can escape away from the sheet. In this subsection, we describe a simple analytic model we use to identify which of the gas clumps identified in the previous section turn into Pop III stars. The main steps in it are as follows:

\begin{enumerate}
    \item We assume that all the clumps follow time evolution similar to the central densest clump as shown in Figure \ref{fig:protostar}. We use it to assign them times of collapse in the future.
    \item Once the time to collapse is reached and the star is formed, it starts to emit ionizing radiation creating a spherical ionized bubble around it. The radius of the ionizing region grows with a parametric shock speed, assumed to be 30 km/s.\footnote{At the high densities in the sheet, the Stromgren sphere is always filled at small radius so the ionization region expands with the speed of the induced shock.} The ionized bubble keeps growing until the star dies and remains of that size afterwards.
    \item At the time of collapse for a given clump, if it happens to be in the ionized region of one of the stars previously formed, the clump does not turn into a star.
\end{enumerate}

\begin{figure*}
    \centering
    \includegraphics[width=\textwidth]{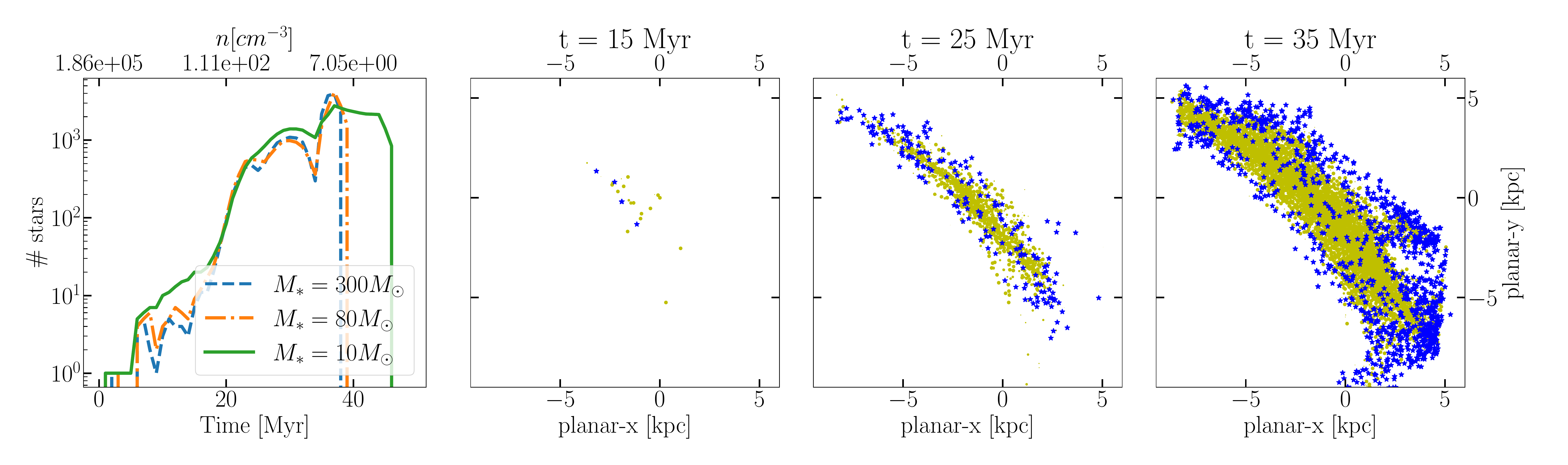}
    \caption{The plot on the left shows the number of stars alive as a function of time calculated using the simple semi-analytic model described in Section \ref{sec:discussion}. The blue (dashed), orange (dot dashed), and green (solid) lines correspond to the stellar mass of 300, 80 and 10 \si{M_{\odot}} which correspond to lifetimes of 2, 3, and 10 Myr, respectively \citep{Schaerer02}. The three panels on the right show the spatial distribution of the stars and HII regions for the case for $M_*$ = 80 $M_\odot$, after 15, 25, and 35 Myr respectively. The chrome yellow shows the ionized region where new stars do not form. The blue symbols denote the stars formed in the past 1 Myr. We can see that as time progresses, the ionized region in the center grows and star formation continues in the outer regions of the sheet.}
    \label{fig:number_stars}
\end{figure*}

Figure \ref{fig:number_stars} shows the number of stars alive as a function of time and location. The left panel shows the number of stars alive as a function of time assuming three different masses of Pop III stars (10, 80, and 300 \si{M_{\odot}}) that correspond to lifetimes of 10, 3, and 2 Myr respectively. 
The difference in the stellar mass does not make a significant difference in the number of stars formed and all of the models predict a burst of Pop III stars with numbers of a few times $10^3$ from 20-40 Myr after the first star forms.
This is mainly because the star-formation starts near the center of the sheet and extends in rings away from the center as the ionized bubbles grow. The peak depends on the large number of gas clumps turning into stars in the outer rings of the sheet. 
The three panels on the right show the growth of the ionized regions (yellow) and the newly formed stars (blue) in the last 1 Myr at 15, 25, and 35 Myr respectively. The star formation starts near the center of the sheet in the plane. As the ionized regions from the stars grow in the central region, star formation continues away from the center in the plane. 

In this simple model, we have assumed that the clump structure in the sheet is not significantly changing with time during the timescale of 40-50 Myr. This approximation is justified as the dynamical time for the sheet using an average density is approximately 120 Myr.
We have also assumed that the growth of the ionized bubble is driven by a $d$-type ionization front shock moving with a fixed velocity of 30 km/s as long as the star is alive. We have also assumed that the ionization and supernovae feedback bubbles stop growing when they reach the edge of the sheet when they encounter the low pressure gas and escape away from the sheet, as shown in the schematic diagram in Figure~\ref{fig:schematic}.

\subsection{Prospects for observations}
The presence of a strong burst of Pop III stars in the pancake is a potential smoking gun signature for the fuzzy dark matter model which can be detected with the newly launched James Webb Space Telescope (JWST). As shown in Figure~\ref{fig:number_stars}, we expect this Pop III galaxy to have a stellar mass of $10^5 -$ \SI{e6}{M_{\odot}} for approximately 20 Myrs. We use \cite{JWST} to make predictions for observational prospects of these objects. \cite{JWST} use a spectral synthesis model \textsc{Yggdrasil} to predict the photometric signatures of Pop III galaxies. They predict that with maximal nebular emission, a Pop III galaxy of mass \SI{4e4}{M_{\odot}} (\SI{2e5}{M_{\odot}}) could be detected with the JWST at $z = 6$ ($z = 10$) for an exposure of 100 hours using NIRCam, assuming a starburst age of 10 Myr. Assuming the typical stellar mass of \SI{80}{M_{\odot}} (orange dot-dashed curve in the left panel of Figure \ref{fig:number_stars}), the simulated galaxy would have a Pop III stellar mass of \SI{8e4}{M_{\odot}}. Thus such a galaxy forming at $z = 6$ can be detected with the JWST.

In the standard model of $\Lambda$CDM, Pop III stars typically form in minihalos of mass $10^5 - $ \SI{e7}{M_{\odot}} with star-formation efficieny $f_* \sim 10^{-4}$ resulting in a total stellar mass of the order of \SI{100}{M_\odot} \citep[e.g.][]{Skinner20}. In cases where Pop III star formation is suppressed in presence of strong Lyman-Werner and ionizing radiation, the star formation can be delayed until the halo becomes large enough to cool via atomic H transitions, when it has a large reservoir of gas. \cite{Kulkarni19} found that Pop III galaxies/clusters forming in massive halos at lower redshift in the presence of a strong ionizing radiation, only have stellar mass of the order of a few times \SI{e3}{M_{\odot}} before the star formation transitions to Pop II. This difference from the FDM is primarily because of the sheet geometry that results in a large number of clumps turning into stars with minimal feedback as described in previous sections. Therefore it is extremely unlikely to form a Pop III galaxy of mass \SI{e5}{M_{\odot}} or more in a $\Lambda$CDM cosmology, thus their presence could be a smoking gun signature for the fuzzy dark matter or alternative dark matter models with small-scale cutoff.

Even though we simulated a Pop III galaxy forming at $z \sim 10$ here, we expect to see similar properties at lower redshift. To estimate the number densities of such Pop III starburst galaxies, we compute the number density of \SI{e10}{M_{\odot}} dark matter halos from the fuzzy dark matter halo mass function using \cite{Kulkarni22_fdm}, which matches well with the estimates from the numerical simulations in \cite{May22} at the relevant scales, with the caveat that the number density of the collapsed halos may not accurately represent the number densities for the Zel'dovich pancakes. At $z = 6$, the comoving number density of these halos correspond to $\sim \SI{e-2}{Mpc^{-3}}$. 
The NIRCam fieldview (2.2 arcmin) at $z = 6$ and using a depth of $\Delta z = 1$ (from $z = 6$ to $z = 7$), corresponding to a comoving volume of \SI{230}{Mpc^3}. Therefore, there should be 2-3 such Pop III starburst galaxies in an FDM cosmology in each NIRCam field of view at $z = 6$.

\section{Summary and conclusion}
\label{sec: conclusions}
Fuzzy dark matter is a promising proposed alternative to $\Lambda$CDM. As small-scale structure is suppressed in an FDM cosmology, the formation of the first stars and galaxies would have delayed formation histories, which could be used to strongly constrain the FDM particle mass. In this letter, we present simulations of the formation of the first galaxies in FDM where we evolve the FDM density accurately by solving the Schr{\"o}dinger-Poisson equations and follow the proto-stellar collapse using very high resolution simulations appropriate for modeling Pop III stars for the first time. We find two novel results: first, the large-scale collapse results in a very thin and flat gas “pancake”; second, despite the very different cosmology, this pancake fragments until it forms protostellar objects, each of which are very similar to those found in CDM minihalos (but much greater in number). Combined, these results indicate that the first generation of stars in FDM cosmologies are also likely to be massive and, because of the sheet morphology, do not self-regulate, resulting in a very massive Pop III starburst. We estimate the total number of first stars forming in this extended structure to be $10^4$ over 20 Myr using a simple model to account for the ionizing feedback from the stars, which should be observable with JWST. These predictions provide a potential smoking gun signature of FDM or similar dark matter candidates.

\begin{acknowledgments}
We thank Jeremiah Ostriker, Zoltan Haiman, and Lam Hui for useful discussions. EV is supported by NSF grant AST-2009309 and NASA grant 80NSSC22K0629.
GLB acknowledges support from the NSF (AST-2108470, XSEDE grant MCA06N030), a NASA TCAN award 80NSSC21K1053, and the Simons Foundation.
The simulations were carried out at Frontera supercomputer of the Texas Advanced Computing Center. XL is supported by the Natural 
Sciences and Engineering Research Council of Canada (NSERC), funding reference \#CITA 490888-16 and the Jeffrey L. Bishop Fellowship.
\end{acknowledgments}

%

\vspace{5mm}


\software{\textsc{Enzo} \citep{Enzo, Enzo_joss}, \textsc{music} \citep{Music}, \textsc{yt} \citep{yt}, \textsc{unyt} \citep{unyt}.
          }




\bibliography{mihir_pop3}{}
\bibliographystyle{aasjournal}



\end{document}